\documentclass[preprint,prd,showpacs,amsmath,amsfonts,amssymb,nofootinbib]{revtex4}
\usepackage[a4paper,lmargin=1in,rmargin=1in]{geometry}
\usepackage{amsmath}
\usepackage{amsfonts}
\usepackage{amsbsy}
\usepackage{amsthm}
\usepackage{accents}
\usepackage{tensor}
\usepackage{enumitem}

\theoremstyle{definition}

\DeclareMathOperator{\End}{End}

\DeclareMathOperator{\tr}{tr}
\DeclareMathOperator{\Tr}{Tr}
\DeclareMathOperator{\vol}{vol}

\newcommand{\rmi}{\mathrm{i}}
\newcommand{\rmd}{\mathrm{d}}

\newcommand{\symped}[1]{\accentset{S}{#1}}
\newcommand{\sympman}{\mathcal{M}}
\newcommand{\bund}{\mathcal{E}}
\newcommand{\ebund}{\End(\bund)}

\newcommand{\lbund}{\mathcal{L}}
\newcommand{\connsymp}{\partial^{S}}
\newcommand{\connbund}{\partial^{\bund}}

\newcommand{\gambund}{\Gamma^{\bund}}

\newcommand{\curvbund}{R^{\bund}}
\newcommand{\curvsymp}{\symped{R}}
\newcommand{\sympvol}{\vol_{S}}
\newcommand{\metvol}{\vol_{M}}

\newcommand{\frmbr}[1]{\langle #1 \rangle}
\newcommand{\ricciend}{\underline{R}}
\newcommand{\riemannend}{\underline{\underline{R}}}
\newcommand{\modricciend}{\underline{\breve{R}}}
\newcommand{\modriemannend}{\underline{\underline{\breve{R}}}}
\newcommand{\tenr}[1]{\tensor{R}{#1}}
\newcommand{\tenx}[1]{\tensor{X}{#1}}
\newcommand{\teny}[1]{\tensor{Y}{#1}}
\newcommand{\tenth}[1]{\tensor{\theta}{#1}}
\newcommand{\tent}[1]{\tensor{\Theta}{#1}}
\newcommand{\tenrl}[1]{\tensor{\undertilde{R}}{#1}}
\newcommand{\tenmodrl}[1]{\tensor{\undertilde{\breve{R}}}{#1}}
\newcommand{\tentor}[1]{\tensor{Q}{#1}}
\newcommand{\zeroed}[1]{\accentset{\mathit{(0)}}{#1}}
\newcommand{\oned}[1]{\accentset{\mathit{(1)}}{#1}}
\newcommand{\twoed}[1]{\accentset{\mathit{(2)}}{#1}}
\newcommand{\teng}[1]{\tensor{\Gamma}{#1}}
\newcommand{\tgzero}[1]{\tensor{\zeroed{\Gamma}}{#1}}
\newcommand{\tgone}[1]{\tensor{\oned{\Gamma}}{#1}}
\newcommand{\tgtwo}[1]{\tensor{\twoed{\Gamma}}{#1}}
\newcommand{\trzero}[1]{\tensor{\zeroed{R}}{#1}}
\newcommand{\trone}[1]{\tensor{\oned{R}}{#1}}
\newcommand{\trtwo}[1]{\tensor{\twoed{R}}{#1}}
\newcommand{\tenxzer}[1]{\tensor{\zeroed{X}}{#1}}
\newcommand{\tenyzer}[1]{\tensor{\zeroed{Y}}{#1}}
\newcommand{\ttorzero}[1]{\tensor{\zeroed{Q}}{#1}}
\newcommand{\ttorone}[1]{\tensor{\oned{Q}}{#1}}
\newcommand{\ttortwo}[1]{\tensor{\twoed{Q}}{#1}}

\begin{document}
\title{On some models of geometric \\ noncommutative general relativity}
\preprint{arXiv:1011.0165}
\author{Micha{\l} Dobrski}
 \email{michal.dobrski@p.lodz.pl}
 \affiliation{
Centre of Mathematics and Physics
\\
Technical University of {\L}\'od\'z,
\\
Al.~Politechniki 11, 90-924 {\L}\'od\'z, Poland}
\date{\today}
\begin{abstract}Using Fedosov theory of deformation quantization of endomorphism bundle we construct several models of pure geometric, deformed vacuum gravity, corresponding to arbitrary symplectic noncommutativity tensor. Deformations of Einstein-Hilbert and Palatini actions are investigated. Coordinate covariant field equations are derived up to the second order of the deformation parameter. For some models they are solved and explicit corrections to an arbitrary Ricci-flat metric are pointed out. The relation to the theory of Seiberg-Witten map is also studied and the correspondence to the spacetime noncommutativity described by Fedosov $*$-product of functions is explained.
\end{abstract}
\maketitle

\section{Introduction}
The present paper is dedicated to a study of some possible global and geometric models of relativity on noncommutative spacetimes within the framework of Fedosov deformation quantization of endomorphism bundle. The motivation for such investigation originates in the conviction that whatever ``noncommutative gravity'' would be, it should preserve the basic symmetry of the classical theory -- the full diffeomorphism invariance. Presented analysis aims at showing that Fedosov quantization of endomorphism bundle can serve as a tool for building geometric field theories on noncommutative spacetimes.

The general strategy we are going to adopt can be summarized in the following steps.
\begin{enumerate}[label=\textit{\arabic*)}]
\item Take some symplectic manifold and an action on it which leads to the general relativity.
\item Rewrite the action by representing Lagrangian as a product of endomorphisms of some bundle. 
\item Replace the product of endomorphisms by Fedosov $*$-product of endomorphisms.
\item Replace the integral by Fedosov trace functional.
\item Do the variations to obtain field equations.
\item Observe that steps \emph{3} and \emph{4} together with results of \cite{dobrski2} induce that the theory is locally equivalent to the theory with Seiberg-Witten map applied on endomorphisms.
\end{enumerate}

There is vast literature concerning construction of noncommutative gravity by means of Moyal product and Seiberg-Witten map. Hence, one can point out series of works \cite{calmet1,calmet2,mukherjee} based on combination of infinitesimal $so(3,1)$ gauging with infinitesimal coordinate transformations, preserving (at first order of deformation) constant deformation parameter $\theta^{ij}$. Another approach is given by \cite{chamseddine1,chamseddine2, cardella, chaichian, mukherjee0A}, where $SO(4,1)$ (or $U(2,2)$) symmetry is investigated. In such setting gauge potential carries information about both tetrad field and the usual $SO(3,1)$-connection. The standard gravity is recovered by the procedure of contraction of the gauge group. There are also investigations based on some variants of $SL(2,\mathbb{C})$ symmetry \cite{chamseddine3, garciacompean}. The common feature of all of these approaches are vanishing first order corrections to the field equations. On the other hand, the common issue is the lack of diffeomorphism invariance\footnote{We are going to distinguish \emph{passive} and \emph{active} diffeomorphism invariance (compare e.g. \cite{rovelli, heinicke}). Here we mean noninvariance in both above senses.}. More general types of noncommutativity were also studied -- eg. Lie algebraic one \cite{mukherjee2} or given by Kontsevich theory \cite{miaozhangsl2c,miaozhangu22}. In \cite{marculescu} theories based on Moyal product and Seiberg-Witten map were geometrized. The resulting structure is invariant under passive  diffeomorphisms, but at the price of nonassociativity of the corresponding $*$-product.
One should also mention some other approaches to noncommutative gravity related somehow to $*$-products and Seiberg-Witten map. These are \cite{aschieri1,aschieri2,aschieri3,aschieri4,meyer,compean2}, where the method of Lie algebra twisting has been used to represent deformation of diffeomorphism symmetry. One of the remarkable results of \cite{aschieri3} is the construction of an action which is geometric (i.e. described by globally defined 4-form) being simultaneously invariant under deformed diffeomorphism symmetry. Finally there are investigations which are strictly related to some particular models emerging in the context of the string theory, e.g. \cite{steinacker1,steinacker2}. 

The paper is organized as follows. In section 2 a brief overview of results of Fedosov construction is presented, and also some further conventions are fixed. In sections 3 and 4 deformations of Einstein-Hilbert and Palatini actions are investigated. Fifth section is devoted to analysis of interrelation between presented models and the theory of Seiberg-Witten map. We also clarify, how obtained results are related to the noncommutativity of the spacetime described by Fedosov $*$-product of functions. Finally some concluding remarks (section 6) are given.
\section{Preliminaries}
\subsection{Fedosov construction}
The main tool used in this paper is Fedosov construction of deformation quantization of endomorphism bundle formulated in \cite{fedosov}. 
We are not going to concern technical or ``internal'' details of this theory (which are interesting and beautiful on their own) but rather to make use of some of its particular results. Interested reader may find short exposition of Fedosov construction in its simplest, suitable for present pourposes form in \cite{dobrski2}. Further geometric and algebraic interpretations are provided by \cite{emmrwein,farkas}. Some other analysis and examples can be found in \cite{tosiek,tosiek2,tosiek3}. Thus, we limit ourselves to the very brief, notation-fixing description of Fedosov $*$-product.

The starting point is given by Fedosov manifold $(\sympman,\omega,\connsymp)$ i.e. $2n$-dimensional symplectic manifold $(\sympman,\omega)$ with some fixed symplectic (torsionless and preserving $\omega$) connection $\connsymp$ \cite{gelfand,bielgutt}. The corresponding Poisson tensor (given by the inverse of $\omega_{ij}$) is going to be denoted as $\Lambda^{ij}$. These data generate\footnote{Precisely, one has also to fix curvature and normalizing condition for Abelian connection generating Fedosov $*$-product \cite{fedosov}. Within this paper, standard normalization $\mu \equiv 0$ and curvature $\Omega=-\omega$ are used.} global, geometric and associative deformation of product of functions on $\sympman$. Its explicit form can be computed up to arbitrary power of deformation parameter $h$ (which has nothing to do with Planck constant in our context) by means of Fedosov's recursive techniques. 

For the vector bundle $\bund$ over $\sympman$, equipped with a connection $\connbund$ one can construct global, geometric and associative deformation of product of sections of $\ebund$. Locally it can be understood as a deformation of product of matrices. Denoting by $\partial=\connsymp\otimes 1+ 1\otimes \connbund$ the connection in $T\sympman\otimes\bund$ (and by the same symbol its natural extension to any other tensor product of $T\sympman$, $T^*\sympman$, $\bund$ and $\bund^*$) one may calculate that for arbitrary two sections $F,G \in C^{\infty}(\ebund)[[h]]$ the Fedosov $*$-product is given up to $h^2$ by the expression
\begin{multline}
\label{fedosov_endstar}
F*G=FG-\frac{\rmi h}{2} \Lambda^{a b} \partial_a F \partial_b G+\\
-\frac{h^2}{8}\Lambda^{ab}\Lambda^{cd}\Big(\{\partial_b F,\curvbund_{ac}\}\partial_d G + \partial_b F \{\curvbund_{ac},\partial_d G\} +\partial_{(a} \partial_{c)} F \partial_{(b} \partial_{d)} G\Big) + O(h^3)
\end{multline}
where $\curvbund_{ab}=\frac{\partial}{\partial x^a}\gambund_b-\frac{\partial}{\partial x^b}\gambund_a+[\gambund_a,\gambund_b]$ (for $\connbund_i=\frac{\partial}{\partial x^i}+\gambund_i$) is the curvature of $\connbund$, and $\{\cdot\,,\cdot\}$ stands for the anticommutator. It is clear that in above formula usual product of endomorphisms (noncommutative from the beginning) has been used. For the special case of flat $\connbund$ and the local frame with $\gambund \equiv 0$, the Fedosov $*$-product of endomorphisms becomes product of matrices with commutative multiplication of entries replaced by noncommutative Fedosov product of functions. 
Such product of matrices is going to be denoted as $*_S$. (The same symbol will be used for the Fedosov product of functions).
If additionally $\connsymp$ is flat and we work in local Darboux coordinates for which coefficients of $\connsymp$ vanish, then Fedosov product of functions becomes Moyal product $*_T$. Thus, in such special case, we are dealing with multiplication used in \cite{seibwitt} for the description of deformed gauge transformations.

The object which needs some more attention is Fedosov trace functional (\cite{fedosov} section 5.6). Given some Fedosov product $*$ one is able to construct trace functional $\tr_*$ taking values in $\mathbb{C}[[h]]$ and acting on compactly supported sections belonging to $C^{\infty}(\ebund)[[h]]$, with the property
\begin{equation}
\label{tracecommutation}
\tr_*(F * G)=\tr_*(G * F).
\end{equation}
If one requires additionally, that for arbitrary (global or local) isomorphism $M$ between $*$-algebras with products $*_1$ and $*_2$ (i.e. for $M$ fulfilling $M(F*_1G)=M(F)*_2M(G)$) the relation
\begin{equation}
\label{traceinvar}
\tr_{*_1}(F)=\tr_{*_2}(M(F))
\end{equation}
holds, then it follows that the trace functional is unique up to a constant normalizing factor. The proof of this fact relies on the observation that for the Moyal product $*_T$ the trace is given by
\begin{equation}
\label{tracemoyal}
\tr_{*_T}(F)= \mathrm{const} \int_{\mathbb{R}^{2n}} \Tr(F) \frac{\omega^n}{n!}
\end{equation} 
where $\Tr$ stands for the trace of a matrix, and on possibility of representing $\tr_*$ in terms of traces on Moyal algebras by a partition of unity $\{\rho_i\}$ and a compatible set of local isomorphisms $\{M_i\}$ between $*$ and the Moyal product. It turns out that $\tr_*$ is independent of particular choice of $\{\rho_i\}$ and $\{A_i\}$. Unlike convention of \cite{fedosov}, we fix the normalizing constant to be equal $1$. Construction presented in \cite{fedosov} enables calculation of explicit form of $\tr_*$. Up to $h^2$ it reads\footnote{The computation leading to (\ref{fedosov_trace}) is quite laborious as one has to deal with connection coefficients which in final step massively cancel and the remaining terms can be grouped to yield tensorial expressions. Large parts of this work has been performed with the significant use of xAct tensor manipulation package \cite{xact}.} 
\begin{equation}
\label{fedosov_trace}
\tr_* (F)=\int_{\sympman} \Tr\Bigg(F + \frac{\rmi h }{2}\Lambda^{a b}\curvbund_{a b} F 
 +h^2 \bigg(-\frac{3}{8} \Lambda^{[a b} \Lambda^{c d]}\curvbund_{a b} \curvbund_{c d} +s_2 1\bigg)F + O(h^3) \Bigg) \frac{\omega^n}{n!}
\end {equation}
where $1$ is the identity endomorphism and the scalar\footnote{Index $_2$ corresponds to the presence $s_2$ at $h^2$. Such defined scalar is a symplectic part of what is called \emph{trace density} in \cite{fedosov}. With similar conventions $s_1=0$.}
\begin{equation}
\label{s2_def}
s_2=\frac{1}{64} \Lambda^{[a b} \Lambda^{c d]}\tensor{\curvsymp}{^k_{lab}}\tensor{\curvsymp}{^l_{kcd}}  + \frac{1}{48}\Lambda^{ab}\Lambda^{cd}\connsymp_e \connsymp_a \tensor{\curvsymp}{^e_{bcd}}
\end{equation}
has been introduced for sake of simplicity of further notations. In above formula $\tensor{\curvsymp}{^i_{jab}}$ stands for the curvature tensor of $\connsymp$. It is useful to write down explicit form of $\tr_*(F*G)$. Substitution of (\ref{fedosov_endstar}) into (\ref{fedosov_trace}) after some manipulations yields\footnote{The formula has been rearranged to explicitly exhibit symmetry $\tr_*(F*G)=\tr_*(G*F)$. For the term at $h$ this can be done quickly using integration by parts and definition of $\curvbund$. For terms at $h^2$ one can proceed in a following manner. \emph{1)} Take what appears at $h^2$ after simple substitution of (\ref{fedosov_endstar}) into (\ref{fedosov_trace}). Let it call $h^2Q(F,G)$. \emph{2)}~Rewrite it as $h^2/2(Q(F,G)+Q(G,F))+h^2/2(Q(F,G)-Q(G,F))$. Drop the antisymmetric part. \emph{3)} Check that discarded terms are indeed equal to zero (as it should be, by the construction). When integrated by parts, the terms with single covariant derivative of $\curvbund$ vanish in virtue of Bianchi identity, while the ones with double $\partial$ can be replaced by $\curvbund$ and in turn, sum up with remaining terms to give 0. Such calculation can be treated as an additional verification of the formula $(\ref{fedosov_trace})$.}
\begin{multline}
\tr_* (F*G)
=\int_{\sympman} \Tr\Bigg(FG+\frac{\rmi h}{4}\Lambda^{a b}\curvbund_{a b} \{F,G\}
+h^2 \Bigg[ s_2 FG
+\frac{1}{8}\Lambda^{a b}\Lambda^{c d}\bigg(\curvbund_{a b}[\partial_c F, \partial_d G]+\\
- \partial_{(a} \partial_{c)} F \partial_{(b} \partial_{d)} G
-\frac{3}{2} \curvbund_{[a b} \curvbund_{c d]}\{F,G\} \bigg)
\Bigg] + O(h^3) \Bigg) \frac{\omega^n}{n!}
\end{multline}
\subsection{Some further conventions}
The important problem related to the programme presented in the introduction is the incompatibility of the volume forms. In (\ref{fedosov_trace}) the symplectic volume form $\sympvol=\frac{\omega^n}{n!}$ must be used,
and in general relativity the metric one $\metvol=\sqrt{-g} \rmd x^1\wedge\dots\wedge \rmd x^{2n}$ more or less explicitly appears. Since the two must be proportional one can write $\metvol=v \sympvol$ defining the function $v:\sympman \to \mathbb{R}$ this way. The above  mentioned incompatibility should be handled somehow, and in what follows two possible approaches are investigated. First, one can simply rescale one of the endomorphisms by multiplying them by $v$. Thus, let us fix the convention that $\breve{F}=vF$. The other option is given by introducing endomorphism $V=v1$ which multiplies endomorphism under the action. Both methods are completely equivalent at the undeformed level, but become different after deformation. 

Let us also point out the following issue concerning the tangent bundle $T\sympman$. In presented models it appears in two distinct roles. First as a "component" of bundle $\bund$, and then as an object which carries information about the symplectic structure and the covariant derivations producing  quantization formalism. This distinction becomes important when applying covariant derivation to tensors involving indices from both copies of $T\sympman$. The $\partial$ connection acts in such case by means of $\connbund$ (which is going to be chosen as a metric connection) and the symplectic connection $\connsymp$ respectively. Thus, one needs some way of ``marking'' indices which should be differentiated by $\connbund$ and we are going to put prime at them (e.g. $\tensor{\curvbund{}}{^{a'}_{b'lm}}$). The ambiguity may be also postponed by using index-free notation for endomorphisms, and this approach is also used. Finally, the primes are omitted in the field equations, as they are no longer needed and may tend to obscure the result. 

Finally, let us mention that all indices in subsequent sections are manipulated by means of corresponding metric tensors. (With exception of relations (\ref{eh_wab_seh1a})--(\ref{eh_wab_seh2b}), (\ref{sol-g2-seh1a}) and (\ref{sol-g2-seh1b}) where the undeformed part of metric is used. This is also recalled within the text). These metric tensors are $g_{ab}$ and also $\eta_{AB}$ for the case of deformation of Palatini action. To avoid ambiguities (or even inconsistencies), we abandon convention of using symplectic form for raising or lowering indices. All formulae taken from Fedosov theory are rewritten in such manner, that they do not involve manipulation of indices by means of symplectic form.
\section{Einstein-Hilbert action}
Now, let us analyze some possible applications of Fedosov theory in the general relativity on noncommutative spacetime. We are going to proceed using programme sketched in the introduction and to assume, that symplectic form $\omega$ and compatible symplectic connection $\connsymp$ are fixed.
First, let us focus on the Einstein-Hilbert action. Thus, there is a metric $g_{ab}$ with determinant $g$, its torsionless Levi-Civita connection $\nabla$, Riemann curvature tensor $\tensor{R}{^{a'}_{b'cd}}$ (also used with all indices primed\footnote{As we use exactly the same frame (e.g. coordinate one) for both primed and unprimed indices we can consistently define primed tensors from unprimed ones and vice versa. The prime is used only as a marking for covariant derivation $\partial$.} $\tensor{R}{^{a'}_{b'c'd'}}$), Ricci tensor $R_{a'b'}=\tensor{R}{^{c'}_{a'c'b'}}$ and Ricci scalar $R$. Field equations are going to be derived by the variation of the metric.

Let us introduce the notation $\tensor{\ricciend}{^{a'}_{b'}}=\tensor{R}{^{a'}_{b'}}$ and $\tensor{\riemannend}{^{a'b'}_{c'd'}}=\tensor{R}{^{a'b'}_{c'd'}}$. (This becomes convenient when distinguishing between endomorphisms $\riemannend$, $\ricciend$ and the scalar $R$).  Also, let
\begin{equation*}
\begin{split}
Y^{ijk'l'}=&\Lambda^{[ij} \Lambda^{ab]}\tensor{R}{^{k'l'}_{ab}}\\
X^{i'j'k'l'}=&\Lambda^{[a b} \Lambda^{c d]} \tenr{^{i'j'}_{ab}}\tenr{^{k'l'}_{cd}} =\tensor{R}{^{i'j'}_{ab}}\tensor{Y}{^{abk'l'}}\\
Z=&\frac{1}{\sqrt{-g}}\Lambda^{ij}\Lambda^{kl}\connsymp_i \connsymp_k \connsymp_j \connsymp_l \sqrt{-g}
\end{split}
\end{equation*}
\subsection{Deformed actions and field equations}
\subsubsection{$\modricciend$ as an endomorphism of $T\sympman$}
The Einstein-Hilbert action can be quickly rewritten as
\begin{equation*}
\mathcal{S}_{EH_{1A}}=\int_{\sympman}\Tr\modricciend\,\frac{\omega^n}{n!}
\end{equation*}
Thus, we are going to treat rescaled Ricci tensor $\tensor{\modricciend}{^{a'}_{b'}}=v\tensor{\ricciend}{^{a'}_{b'}}$ as an endomorphism of $\bund=T\sympman$. In order to define $*$-product of endomorphisms one needs some connection in $\bund$. Let us fix $\connbund=\nabla$ and consequently $\curvbund$ is given by the Riemann tensor. The corresponding $*$-product is going to be denoted by $*_{EH_1}$. Under these assumptions the deformed action is given by
\begin{equation*}
\begin{split}
\widehat{\mathcal{S}}_{EH_{1A}}=\tr_{*_{EH_1}}(\modricciend)=&\int_{\sympman} \Tr\Bigg(\modricciend  
 +h^2 \bigg(-\frac{3}{8} \Lambda^{[a b} \Lambda^{c d]}\curvbund_{a b} \curvbund_{c d} +s_2\bigg)\modricciend + O(h^3) \Bigg) \frac{\omega^n}{n!}\\
=&\int_{\sympman} \bigg(R
-\frac{3}{8}h^2 \tensor{X}{^{k'}_{l'}^{l'}_{m'}}\tensor{R}{^{m'}_{k'}}
  +h^2 s_2 R + O(h^3) \bigg) \metvol
\end{split}
\end{equation*}
Variation of the metric yields the following field equations
\begin{equation*}
\begin{split}
&R^{ab}-\frac{1}{2}g^{ab}R+h^2\Bigg[\frac{3}{8}\bigg(
-\tenr{^{(a}_k}\tenx{_l^{b)}^k^l}
+\frac{1}{2}\tenr{^k_l}\tenx{^l_m^m_k}g^{ab}+
\nabla_k \nabla^{(a} \tenx{^{b)}_l^{lk}}
-\frac{1}{2}\nabla_l \nabla^l \tenx{^a_k^{kb}}\\
&-\frac{1}{2}g^{ab}\nabla_k \nabla_l \tenx{^k_m^{ml}}
-2\nabla_k \nabla^l\left(\tenr{^{(a}_m}\teny{_l^{b)mk}}\right)
+2\nabla_k \nabla_l\left(\tenr{^{km}}\teny{^{l(a}_m^{b)}} \right)
\bigg)
-\frac{1}{2}g^{ab}Rs_2\\
&+R^{ab}s_2
+g^{ab}\nabla_l \nabla^l s_2-\nabla^a \nabla^b s_2
\Bigg]+O(h^3)=0
\end{split}
\end{equation*}
\subsubsection{$\ricciend$ and $V$ as endomorphisms of $T\sympman$}
\label{RVendTM}
Now, keeping unmodified $*$-product structure given by $*_{EH_1}$, we are going to investigate another possibility of forcing correct volume form at $h=0$. The Einstein-Hilbert action written as
\begin{equation*}
\mathcal{S}_{EH_{1B}}=\int_{\sympman}\Tr\ricciend V\,\frac{\omega^n}{n!}
\end{equation*}
may be deformed into\footnote{The term with $\partial_{(a} \partial_{c)} \ricciend  \partial_{(b} \partial_{d)} V$ is integrated by parts twice, then the covariant derivatives are commuted with the trace and the torsionless property of $\connsymp$ is used to get rid of symmetrizations.}
\begin{equation*}
\begin{split}
&\widehat{\mathcal{S}}_{EH_{1B}}=\tr_{*_{EH_1}}(\ricciend *_{EH_1} V)\\
&=\int_{\sympman} \Tr\Bigg(\ricciend V
+h^2 \bigg(-\frac{1}{8}\Lambda^{a b}\Lambda^{c d}\Big(  
 \partial_{(a} \partial_{c)} \ricciend  \partial_{(b} \partial_{d)} V
+3 \curvbund_{[a b} \curvbund_{c d]}\ricciend V \Big)+s_2 \ricciend V
\bigg) + O(h^3) \Bigg) \frac{\omega^n}{n!}\\
&=\int_{\sympman} \bigg(R
-\frac{3}{8}h^2 \tensor{X}{^{k'}_{l'}^{l'}_{m'}}\tensor{R}{^{m'}_{k'}}
-\frac{1}{8}h^2 \Lambda^{a b}\Lambda^{c d} \connsymp_{b}\connsymp_{d}\connsymp_{a}\connsymp_{c} R 
+h^2 s_2 R + O(h^3) \bigg) \metvol
\end{split}
\end{equation*}
Then, the field equations become
\begin{equation*}
\begin{split}
&R^{ab}-\frac{1}{2}g^{ab}R+h^2\Bigg[\frac{3}{8}\bigg(
-\tenr{^{(a}_k}\tenx{_l^{b)}^k^l}
+\frac{1}{2}\tenr{^k_l}\tenx{^l_m^m_k}g^{ab}+
\nabla_k \nabla^{(a} \tenx{^{b)}_l^{lk}}
-\frac{1}{2}\nabla_l \nabla^l \tenx{^a_k^{kb}}\\
&-\frac{1}{2}g^{ab}\nabla_k \nabla_l \tenx{^k_m^{ml}}
-2\nabla_k \nabla^l\left(\tenr{^{(a}_m}\teny{_l^{b)mk}}\right)
+2\nabla_k \nabla_l\left(\tenr{^{km}}\teny{^{l(a}_m^{b)}} \right)
\bigg)
+\frac{1}{8}\bigg(
-\tenr{^{ab}}Z\\
&+\nabla^a \nabla^b Z-g^{ab}\nabla_l\nabla^l Z
+\frac{1}{2}g^{ab}\Lambda^{jk}\Lambda^{lm} \connsymp_{k}\connsymp_{m}\connsymp_{j}\connsymp_{l}R
\bigg)
-\frac{1}{2}g^{ab}Rs_2+R^{ab}s_2 \\
&+g^{ab}\nabla_l \nabla^l s_2-\nabla^a \nabla^b s_2
\Bigg]+O(h^3)=0
\end{split}
\end{equation*}
\subsubsection{$\modriemannend$ as an endomorphism of $T\sympman \otimes T\sympman$}
This time, we start with the action
\begin{equation*}
\mathcal{S}_{EH_{2A}}=\int_{\sympman}\Tr\modriemannend\,\frac{\omega^n}{n!}
\end{equation*}
Here, the rescaled Riemann tensor is treated as an endomorphism of $\bund=T\sympman \otimes T\sympman$ whose action on $l \in T\sympman \otimes T\sympman$ yields $(\modriemannend l)^{a'b'} = v\tensor{R}{^{a'b'}_{c'd'}}l^{c'd'}$. As a connection in $\bund$ we take $\connbund=\nabla \otimes 1+1 \otimes \nabla$. Its curvature is given by $\curvbund_{ab}=R^{\nabla}_{ab}\otimes1+1\otimes R^{\nabla}_{ab}$, with $R^{\nabla}_{ab}$ being the curvature of $\nabla$ treated as an endomorphism of $T\sympman$. Let $*_{EH_2}$ be the corresponding $*$-product. Thus
\begin{equation*}
\begin{split}
\widehat{\mathcal{S}}_{EH_{2A}}=&\tr_{*_{EH_2}}(\modriemannend)=\int_{\sympman} \Tr\Bigg(\modriemannend  
 +h^2 \bigg(-\frac{3}{8} \Lambda^{[a b} \Lambda^{c d]}\curvbund_{a b} \curvbund_{c d} +s_2\bigg)\modriemannend + O(h^3) \Bigg) \frac{\omega^n}{n!}\\
=&\int_{\sympman} \bigg(R
 -\frac{3}{4}h^2 \Big(\tensor{X}{^{k'}_{l'}^{l'}_{m'}}\tensor{R}{^{m'}_{k'}}
+ \tensor{X}{^{k'}_{l'}^{m'}_{p'}}\tensor{R}{^{l'p'}_{k'm'}}\Big) +h^2 s_2 R + O(h^3) \bigg) \metvol
\end{split}
\end{equation*}
yielding
\begin{equation*}
\begin{split}
&R^{ab}-\frac{1}{2}g^{ab}R+h^2\Bigg[\frac{3}{4}\bigg(
-\tenr{^{(a}_k}\tenx{_l^{b)}^k^l}
+\frac{1}{2}\tenr{^k_l}\tenx{^l_m^m_k}g^{ab}+
\nabla_k \nabla^{(a} \tenx{^{b)}_l^{lk}}
-\frac{1}{2}\nabla_l \nabla^l \tenx{^a_k^{kb}}\\
&-\frac{1}{2}g^{ab}\nabla_k \nabla_l \tenx{^k_m^{ml}}
-2\nabla_k \nabla^l\left(\tenr{^{(a}_m}\teny{_l^{b)mk}}\right)
+2\nabla_k \nabla_l\left(\tenr{^{km}}\teny{^{l(a}_m^{b)}} \right)
+\tenr{_{km}^{l(a}}\tenx{^{b)m}^k_l}\\
&+\frac{1}{2}\nabla_k \nabla_l \tenx{^{k(ab)l}}
+2 \nabla_k \nabla_l\left( \tenr{^{mjk(a}}\teny{^{b)l}_{mj}} \right)
+\frac{1}{2}\tenr{^{lm}_{jk}}\tenx{^j_l^k_m}g^{ab} 
\bigg)
-\frac{1}{2}g^{ab}Rs_2+R^{ab}s_2 \\
&+g^{ab}\nabla_l \nabla^l s_2-\nabla^a \nabla^b s_2
\Bigg]+O(h^3)=0
\end{split}
\end{equation*}
\subsubsection{$\riemannend$ and $V$ as endomorphisms of $T\sympman \otimes T\sympman$}
Analogously to section \ref{RVendTM}, one may keep the product $*_{EH_2}$ unchanged, but rewrite the action using $V$
\begin{equation*}
\mathcal{S}_{EH_{2B}}=\int_{\sympman}\Tr\riemannend V\,\frac{\omega^n}{n!}
\end{equation*}
After the deformation the action takes form
\begin{equation*}
\begin{split}
&\widehat{\mathcal{S}}_{EH_{2B}}=\tr_{*_{EH_2}}(\riemannend *_{EH_2} V)\\
&=\int_{\sympman} \Tr\Bigg(\riemannend V
+h^2 \bigg(-\frac{1}{8}\Lambda^{a b}\Lambda^{c d}\Big(  
 \partial_{(a} \partial_{c)} \riemannend  \partial_{(b} \partial_{d)} V
+3 \curvbund_{[a b} \curvbund_{c d]}\riemannend V \Big)+s_2 \riemannend V
\bigg) + O(h^3) \Bigg) \frac{\omega^n}{n!}\\
&=\int_{\sympman} \bigg(R
 -\frac{3}{4}h^2 \Big(\tensor{X}{^{k'}_{l'}^{l'}_{m'}}\tensor{R}{^{m'}_{k'}}
+ \tensor{X}{^{k'}_{l'}^{m'}_{p'}}\tensor{R}{^{l'p'}_{k'm'}}\Big) 
-\frac{1}{8}h^2 \Lambda^{a b}\Lambda^{c d} \connsymp_{b}\connsymp_{d}\connsymp_{a}\connsymp_{c} R+h^2 s_2 R\\
&\quad+ O(h^3)\bigg) \metvol  
\end{split}
\end{equation*}
The field equations are given by
\begin{equation*}
\begin{split}
&R^{ab}-\frac{1}{2}g^{ab}R+h^2\Bigg[\frac{3}{4}\bigg(
-\tenr{^{(a}_k}\tenx{_l^{b)}^k^l}
+\frac{1}{2}\tenr{^k_l}\tenx{^l_m^m_k}g^{ab}+
\nabla_k \nabla^{(a} \tenx{^{b)}_l^{lk}}
-\frac{1}{2}\nabla_l \nabla^l \tenx{^a_k^{kb}}\\
&-\frac{1}{2}g^{ab}\nabla_k \nabla_l \tenx{^k_m^{ml}}
-2\nabla_k \nabla^l\left(\tenr{^{(a}_m}\teny{_l^{b)mk}}\right)
+2\nabla_k \nabla_l\left(\tenr{^{km}}\teny{^{l(a}_m^{b)}} \right)
+\tenr{_{km}^{l(a}}\tenx{^{b)m}^k_l}\\
&+\frac{1}{2}\nabla_k \nabla_l \tenx{^{k(ab)l}}
+2 \nabla_k \nabla_l\left( \tenr{^{mjk(a}}\teny{^{b)l}_{mj}} \right)
+\frac{1}{2}\tenr{^{lm}_{jk}}\tenx{^j_l^k_m}g^{ab} 
\bigg)
+\frac{1}{8}\bigg(
-\tenr{^{ab}}Z\\
&+\nabla^a \nabla^b Z-g^{ab}\nabla_l\nabla^l Z
+\frac{1}{2}g^{ab}\Lambda^{jk}\Lambda^{lm} \connsymp_{k}\connsymp_{m}\connsymp_{j}\connsymp_{l}R
\bigg)
-\frac{1}{2}g^{ab}Rs_2+R^{ab}s_2 \\
&+g^{ab}\nabla_l \nabla^l s_2-\nabla^a \nabla^b s_2
\Bigg]+O(h^3)=0
\end{split}
\end{equation*}
\subsection{Structure of deformed theories}
Let us briefly comment formulae obtained in the previous subsection. In all cases the $h^1$ terms in deformed actions have vanished due to $\tensor{R}{^k_{lab}}\tensor{R}{^l_k}=0$. Also, in all deformed Lagrangians one is dealing with $h^2\twoed{\mathcal{L}}_s=h^2 s_2 R$ term, originating in the part of the trace formula (\ref{fedosov_trace}) generated by the curvature of symplectic connection. $\twoed{\mathcal{L}}_s$ gives rise to the field equations by the expression 
\begin{equation*}
-\frac{1}{2}g^{ab}Rs_2+R^{ab}s_2 +g^{ab}\nabla_l \nabla^l s_2-\nabla^a \nabla^b s_2,
\end{equation*}
and describes interaction of the metric $g_{ab}$ with symplectic connection, which defines $s_2$. 

Now, let us write Lagrangians (with respect to $\metvol$) as
\begin{align*}
\widehat{\mathcal{L}}_{EH_{1A}}&=\mathcal{L}_{EH}+h^2\twoed{\mathcal{L}}_{\modricciend}+h^2\twoed{\mathcal{L}}_s +O(h^3),\\
\widehat{\mathcal{L}}_{EH_{1B}}&=\mathcal{L}_{EH}+h^2\twoed{\mathcal{L}}_{\ricciend *_{EH_1} V}+h^2\twoed{\mathcal{L}}_s +O(h^3),\\
\widehat{\mathcal{L}}_{EH_{2A}}&=\mathcal{L}_{EH}+h^2\twoed{\mathcal{L}}_{\modriemannend}+h^2\twoed{\mathcal{L}}_s +O(h^3),\\
\widehat{\mathcal{L}}_{EH_{2B}}&=\mathcal{L}_{EH}+h^2\twoed{\mathcal{L}}_{\riemannend *_{EH_1} V}+h^2\twoed{\mathcal{L}}_s +O(h^3),
\end{align*}
with $\mathcal{L}_{EH}=R$. Hence, $\twoed{\mathcal{L}}_{\modricciend}$, $\twoed{\mathcal{L}}_{\ricciend *_{EH_1} V}$, $\twoed{\mathcal{L}}_{\modriemannend}$ and $\twoed{\mathcal{L}}_{\riemannend *_{EH_1} V}$ represent terms produced by the part of the trace generated by $\curvbund$.
It follows that choosing $\bund=T\sympman$ and $\modricciend$ as the endomorphism yields $\twoed{\mathcal{L}}_{\modricciend}=-\frac{3}{8} \tensor{X}{^{k'}_{l'}^{l'}_{m'}}\tensor{R}{^{m'}_{k'}}$, while taking $\bund=T\sympman \otimes T\sympman$ and $\modriemannend$ produces $\twoed{\mathcal{L}}_{\modriemannend}=-\frac{3}{4} \Big(\tensor{X}{^{k'}_{l'}^{l'}_{m'}}\tensor{R}{^{m'}_{k'}}
+ \tensor{X}{^{k'}_{l'}^{m'}_{p'}}\tensor{R}{^{l'p'}_{k'm'}}\Big)$. Thus one can write the relation $\twoed{\mathcal{L}}_{\modriemannend}=2\twoed{\mathcal{L}}_{\modricciend}-\frac{3}{4}\tensor{X}{^{k'}_{l'}^{m'}_{p'}}\tensor{R}{^{l'p'}_{k'm'}}$. Switching from endomorphism rescaling to multiplication by $V$ 
influences the deformed Lagrangians by $\twoed{\mathcal{L}}_{\ricciend *_{EH_1} V}=\twoed{\mathcal{L}}_{\modricciend}-\frac{1}{8}h^2 \Lambda^{a b}\Lambda^{c d} \connsymp_{b}\connsymp_{d}\connsymp_{a}\connsymp_{c} R$ 
, and $\twoed{\mathcal{L}}_{\riemannend *_{EH_2} V}=\twoed{\mathcal{L}}_{\modriemannend}-\frac{1}{8}h^2 \Lambda^{a b}\Lambda^{c d} \connsymp_{b}\connsymp_{d}\connsymp_{a}\connsymp_{c} R$.

The term $\twoed{\mathcal{L}}_{\modricciend}$ contributes to the field equations by
\begin{multline*}
\frac{3}{8}\bigg(
-\tenr{^{(a}_k}\tenx{_l^{b)}^k^l}
+\frac{1}{2}\tenr{^k_l}\tenx{^l_m^m_k}g^{ab}+
\nabla_k \nabla^{(a} \tenx{^{b)}_l^{lk}}
-\frac{1}{2}\nabla_l \nabla^l \tenx{^a_k^{kb}}\\
-\frac{1}{2}g^{ab}\nabla_k \nabla_l \tenx{^k_m^{ml}}
-2\nabla_k \nabla^l\left(\tenr{^{(a}_m}\teny{_l^{b)mk}}\right)
+2\nabla_k \nabla_l\left(\tenr{^{km}}\teny{^{l(a}_m^{b)}} \right)
\bigg).
\end{multline*}
Analogously, from $-\frac{3}{4}\tensor{X}{^{k'}_{l'}^{m'}_{p'}}\tensor{R}{^{l'p'}_{k'm'}}$ one obtains
\begin{equation*}
\frac{3}{4}\bigg(\tenr{_{km}^{l(a}}\tenx{^{b)m}^k_l}
+\frac{1}{2}\nabla_k \nabla_l \tenx{^{k(ab)l}}
+2 \nabla_k \nabla_l\left( \tenr{^{mjk(a}}\teny{^{b)l}_{mj}} \right)
+\frac{1}{2}\tenr{^{lm}_{jk}}\tenx{^j_l^k_m}g^{ab} 
\bigg).
\end{equation*}
Finally, the expression $-\frac{1}{8}h^2 \Lambda^{a b}\Lambda^{c d} \connsymp_{b}\connsymp_{d}\connsymp_{a}\connsymp_{c} R$ is responsible for 
\begin{equation*}
\frac{1}{8}\bigg(
-\tenr{^{ab}}Z
+\nabla^a \nabla^b Z-g^{ab}\nabla_l\nabla^l Z
+\frac{1}{2}g^{ab}\Lambda^{jk}\Lambda^{lm} \connsymp_{k}\connsymp_{m}\connsymp_{j}\connsymp_{l}R
\bigg),
\end{equation*}
being the second source of terms involving symplectic connection.
\subsection{Corrections to the metric}
In all considered cases field equations are of the form $G_{ab}=W_{ab}+O(h^3)$, where $G_{ab}=R_{ab}-\frac{1}{2}R g_{ab}$ is the Einstein tensor, and the term $W_{ab}$ is of $h^2$ order i.e.  $W_{ab}=h^2 \twoed{W}_{ab}+O(h^3)$. Let us investigate how $W_{ab}$ influences the metric. For this pourpose one can rewrite $g_{ab}$ as a formal power series with respect to $h$
\begin{equation*}
g_{ab}=\zeroed{g}_{ab}+h\oned{g}_{ab}+h^2 \,\twoed{g}_{ab}+\dots
\end{equation*}
Coefficients of Levi-Civita connection corresponding to $g_{ab}$ can be written as
\begin{equation*}
\teng{^a_{bc}}=\tgzero{^a_{bc}}+h\tgone{^a_{bc}}+h^2\,\tgtwo{^a_{bc}}+\dots
\end{equation*}
One can quite easily calculate that
\begin{subequations}
\begin{eqnarray}
\label{eh_gamma_zero}
\tgzero{^a_{bc}} &=& \frac{1}{2}\zeroed{g}^{\;ak}\left( \frac{\partial \zeroed{g}_{kb}}{\partial x^c }+ \frac{\partial \zeroed{g}_{kc}}{\partial x^b } - 
\frac{\partial \zeroed{g}_{bc}}{\partial x^k }\right),\\
\label{eh_gamma_one}
\tgone{^a_{bc}} &=& \frac{1}{2}\zeroed{g}^{\;ak}\left(\zeroed{\nabla}_{c} \oned{g}_{kb} + \zeroed{\nabla}_{b} \oned{g}_{kc} - \zeroed{\nabla}_{k} \oned{g}_{bc}\right),\\
\tgtwo{^a_{bc}} &=& \frac{1}{2}\zeroed{g}^{\;ak}\left(\zeroed{\nabla}_{c} \twoed{g}_{kb} + \zeroed{\nabla}_{b} \twoed{g}_{kc} - \zeroed{\nabla}_{k} \twoed{g}_{bc}\right)
- \zeroed{g}^{\;ak\;} \oned{g}_{kl} \tgone{^{\,l}_{bc}},
\end{eqnarray}
\end{subequations}
where $\zeroed{\nabla}$ denotes Levi-Civita connection of metric $\zeroed{g}_{ab}$. Observe that  $\tgone{^a_{bc}}$ and $\tgtwo{^a_{bc}}$ are tensorial objects. Hence, for the Riemann tensor
\begin{equation*}
\tenr{^a_{bcd}}=\trzero{^a_{bcd}}+h\trone{^a_{bcd}}+h^2\trtwo{^a_{bcd}}+\dots
\end{equation*}
 one obtains
\begin{subequations}
\label{eh_riemann_corr}
\begin{eqnarray}
\trone{^a_{bcd}} &=& 2\zeroed{\nabla}_{[c} \tgone{^a_{d]b}},\\
\trtwo{^a_{bcd}} &=& 2\zeroed{\nabla}_{[c} \tgtwo{^a_{d]b}} + 2\tgone{^a_{k[c}}\tgone{^k_{d]b}},
\end{eqnarray}
\end{subequations}
and $\trzero{^a_{bcd}}$ is the Riemann tensor of metric $\zeroed{g}_{ab}$. Substituting above relations into field equations and analyzing terms at $h^0$ and $h^1$ one calculates that
\begin{subequations}
\label{eh_g_corr}
\begin{gather}
\label{eh_g0_corr}
\trzero{_{ab}} = 0,\\
\label{eh_g1_corr}
\zeroed{g}^{\;kl}\left(\zeroed{\nabla}_k \zeroed{\nabla}_a \oned{g}_{bl} + \zeroed{\nabla}_k \zeroed{\nabla}_b \oned{g}_{al} - \zeroed{\nabla}_k \zeroed{\nabla}_l \oned{g}_{ab} - \zeroed{\nabla}_a \zeroed{\nabla}_b \oned{g}_{kl} \right)=0,
\end{gather}
where $\trzero{_{ab}}$ is zeroth order term in power series expansion of $R_{ab}$ and also Ricci tensor of $\zeroed{g}_{ab}$. For $h^2$ the following relation can be derived
\begin{multline}
\label{eh_g2_corr}
\zeroed{g}^{\;kl}\left(\zeroed{\nabla}_k \zeroed{\nabla}_a \twoed{g}_{bl} + \zeroed{\nabla}_k \zeroed{\nabla}_b \twoed{g}_{al} - \zeroed{\nabla}_k \zeroed{\nabla}_l \twoed{g}_{ab} - \zeroed{\nabla}_a \zeroed{\nabla}_b \twoed{g}_{kl} 
\right)=\\
=2\twoed{W}_{ab}- \frac{1}{n-1}\zeroed{g}_{ab} \twoed{W}
-4\tgone{^k_{l[k}}\tgone{^l_{b]a}}
+4\,\zeroed{g}^{\;rk} \zeroed{\nabla}_{[r} \left(\,\tgone{^l_{b]a}} \oned{g}_{kl} \right),
\end{multline}
where $\twoed{W}=\zeroed{g}^{\;rs} \twoed{W}_{rs}$. The term $\twoed{W}_{ab}$ is given by the following formulae\footnote{In equations (\ref{eh_wab_seh1a}--\ref{eh_wab_seh2b}) indices are manipulated by means of metric $\zeroed{g}_{ab}$.}:
\begin{itemize}[label=--]
\item for $\widehat{\mathcal{S}}_{EH_{1A}}$
\begin{equation}
\label{eh_wab_seh1a}
\begin{split}
\twoed{W}_{ab}=&-\frac{3}{8}\Bigg(
\zeroed{\nabla}_k \zeroed{\nabla}_{(a} \tenxzer{_{b)}_l^{lk}}
-\frac{1}{2}\zeroed{\nabla}_l \zeroed{\nabla}^l \tenxzer{_a_k^k_b}
-\frac{1}{2}\zeroed{g}_{ab}\zeroed{\nabla}_k \zeroed{\nabla}_l \tenxzer{^k_m^{ml}}
\Bigg)\\
&-\zeroed{g}_{ab}\zeroed{\nabla}_l \zeroed{\nabla}^l s_2+\zeroed{\nabla}_a \zeroed{\nabla}_b s_2,
\end{split}
\end{equation}
\item for $\widehat{\mathcal{S}}_{EH_{1B}}$
\begin{equation}
\begin{split}
\twoed{W}_{ab}=&-\frac{3}{8}\Bigg(
\zeroed{\nabla}_k \zeroed{\nabla}_{(a} \tenxzer{_{b)}_l^{lk}}
-\frac{1}{2}\zeroed{\nabla}_l \zeroed{\nabla}^l \tenxzer{_a_k^k_b}
-\frac{1}{2}\zeroed{g}_{ab}\zeroed{\nabla}_k \zeroed{\nabla}_l \tenxzer{^k_m^{ml}}
\Bigg)\\
&-\frac{1}{8}\Bigg(
\zeroed{\nabla}_a \zeroed{\nabla}_b \zeroed{Z}-\zeroed{g}_{ab}\zeroed{\nabla}_l\zeroed{\nabla}^l \zeroed{Z}
\Bigg)
-\zeroed{g}_{ab}\zeroed{\nabla}_l \zeroed{\nabla}^l s_2+\zeroed{\nabla}_a \zeroed{\nabla}_b s_2,
\end{split}
\end{equation}
\item for $\widehat{\mathcal{S}}_{EH_{2A}}$
\begin{equation}
\begin{split}
\twoed{W}_{ab}=&-\frac{3}{4}\Bigg(
\zeroed{\nabla}_k \zeroed{\nabla}_{(a} \tenxzer{_{b)}_l^{lk}}
-\frac{1}{2}\zeroed{\nabla}_l \zeroed{\nabla}^l \tenxzer{_a_k^k_b}
-\frac{1}{2}\zeroed{g}_{ab}\zeroed{\nabla}_k \zeroed{\nabla}_l \tenxzer{^k_m^{ml}}
+\trzero{_{km}^l_{(a}}\tenxzer{_{b)}^m^k_l}\\
&+\frac{1}{2}\zeroed{\nabla}_k \zeroed{\nabla}_l \tenxzer{^{k}_{(ab)}^l}
+2 \zeroed{\nabla}_k \zeroed{\nabla}_l\left( \trzero{^{\,mjk}_{(a}}\tenyzer{_{b)}^l_{mj}} \right)
+\frac{1}{2}\trzero{^{\,lm}_{jk}}\tenxzer{^j_l^k_m}\zeroed{g}_{ab} 
\Bigg)\\
&
-\zeroed{g}_{ab}\zeroed{\nabla}_l \zeroed{\nabla}^l s_2+\zeroed{\nabla}_a \zeroed{\nabla}_b s_2,
\end{split}
\end{equation}
\item for $\widehat{\mathcal{S}}_{EH_{2B}}$
\begin{equation}
\label{eh_wab_seh2b}
\begin{split}
\twoed{W}_{ab}=&-\frac{3}{4}\Bigg(
\zeroed{\nabla}_k \zeroed{\nabla}_{(a} \tenxzer{_{b)}_l^{lk}}
-\frac{1}{2}\zeroed{\nabla}_l \zeroed{\nabla}^l \tenxzer{_a_k^k_b}
-\frac{1}{2}\zeroed{g}_{ab}\zeroed{\nabla}_k \zeroed{\nabla}_l \tenxzer{^k_m^{ml}}
+\trzero{_{km}^l_{(a}}\tenxzer{_{b)}^m^k_l}\\
&+\frac{1}{2}\zeroed{\nabla}_k \zeroed{\nabla}_l \tenxzer{^{k}_{(ab)}^l}
+2 \zeroed{\nabla}_k \zeroed{\nabla}_l\left( \trzero{^{\,mjk}_{(a}}\tenyzer{_{b)}^l_{mj}} \right)
+\frac{1}{2}\trzero{^{\,lm}_{jk}}\tenxzer{^j_l^k_m}\zeroed{g}_{ab} 
\Bigg)\\
&-\frac{1}{8}\Bigg(
\zeroed{\nabla}_a \zeroed{\nabla}_b \zeroed{Z}-\zeroed{g}_{ab}\zeroed{\nabla}_l\zeroed{\nabla}^l \zeroed{Z}
\Bigg)
-\zeroed{g}_{ab}\zeroed{\nabla}_l \zeroed{\nabla}^l s_2+\zeroed{\nabla}_a \zeroed{\nabla}_b s_2,
\end{split}
\end{equation}
\end{itemize}
\end{subequations}
where $\tenxzer{^{ijkl}}$, $\tenyzer{^{ijkl}}$, $\zeroed{Z}$ are zeroth order terms in power series expansion of  $X_{ijkl}$, $Y_{ijkl}$ and $Z$, which can be expressed by means of $\trzero{^i_{jkl}}$ and $\zeroed{g}_{ab}$. 

Thus, in all cases equations which describe deformed metric are of the same structure. At $h^0$ one is dealing with arbitrary Ricci-flat metric $\zeroed{g}_{ab}$. The $h^1$ correction $\oned{g}_{ab}$ can be understood as a classical\footnote{Arguments  leading to (\ref{eh_g1_corr}) are essentially identical to standard calculations concerning small perturbations of classical vacuum relativity, e.g. in \emph{shortwave formalism} (\cite{misthrnwh} \S 35.13).} (undeformed) first order perturbation of $\zeroed{g}_{ab}$, governed by the linear homogeneous equations (\ref{eh_g1_corr}). Noncommutativity appears for the first time at $h^2$. Correction $\twoed{g}_{ab}$ is given by linear, inhomogeneous equations (\ref{eh_g2_corr}). The homogeneous part of (\ref{eh_g2_corr}) is expressed by the same linear operator as that of (\ref{eh_g1_corr}). The inhomogeneous part consists of two groups of terms -- the one describing interaction with first order perturbation $\oned{g}_{ab}$, and the other given by $W_{ab}$, with purely noncommutative origin. Discarding first order classical perturbation by putting $\oned{g}_{ab}=0$, we are able to point out special solution of (\ref{eh_g2_corr}) for actions $\widehat{\mathcal{S}}_{EH_{1A}}$ and  $\widehat{\mathcal{S}}_{EH_{1B}}$. It reads
\begin{equation}
\label{sol-g2-seh1a}
\twoed{g}_{ab}=-\frac{3}{8} \tenxzer{_{ak}^k_b} - \frac{1}{n-1}\left(s_2 -\frac{3}{16} \tenxzer{_{mk}^{km}} \right)\zeroed{g}_{ab}
\end{equation}
for $\widehat{\mathcal{S}}_{EH_{1A}}$, and
\begin{equation}
\label{sol-g2-seh1b}
\twoed{g}_{ab}=-\frac{3}{8} \tenxzer{_{ak}^k_b} - \frac{1}{n-1}\left(s_2 -\frac{1}{8}\zeroed{Z} -\frac{3}{16} \tenxzer{_{mk}^{km}} \right)\zeroed{g}_{ab}
\end{equation}
for $\widehat{\mathcal{S}}_{EH_{1B}}$. (Here, like in (\ref{eh_wab_seh1a}) -- (\ref{eh_wab_seh2b}), indices at $\tenxzer{}$ are manipulated by $\zeroed{g}_{ab}$). 
Let us observe that the difference between above solutions and arbitrary other solution of (\ref{eh_g2_corr}), with $\oned{g}_{ab}=0$, must be a solution of homogeneous variant of (\ref{eh_g2_corr}). Thus, such a difference may be interpreted as a classical perturbation of metric $\zeroed{g}_{ab}$. For this reason one can regard (\ref{sol-g2-seh1a}) and (\ref{sol-g2-seh1b}) as the solutions carrying full information about considered noncommutativity at $h^2$.
\section{Palatini action}
Now, let us switch to the Palatini formalism with the connection and the tetrad field as separate dynamical variables. Thus, one is dealing with the vector bundle $\lbund$ for which  $SO(3,1)$ transformations preserve the canonical form of the Lorentzian metric $\eta_{AB}$. The bundle $\lbund$ is equipped with some metric-compatible connection $\partial^{\lbund}$ . Its local coefficients are denoted as $\tensor{\undertilde{\Gamma}}{^{AB}_i}$ and are antisymmetric in $^{AB}$. The corresponding curvature is given by $\tenrl{^A_{Bij}}$. The bundle $\bund$ is taken to be $\lbund \otimes T\sympman$. The tetrad field $\tenth{^A_{b'}}$ induces the metric $g_{a'b'}=\tenth{^A_{a'}}\eta_{AB}\tenth{^B_{b'}}$ and the metric connection $\nabla$ in $T\sympman$ (not necessarily torsionless). Local coefficients of $\nabla$ can be computed from the expression $\tensor{\Gamma}{^{i'}_{j'k}}
=\tenth{_A^{i'}}\tensor{\undertilde{\Gamma}}{^A_B_k}\tenth{^B_{j'}}+\tenth{_A^{i'}}\frac{\partial}{\partial x^k}\tenth{^A_{j'}}$. The curvature tensors are related by $\tensor{R}{^{i'}_{j'kl}}=\tenth{_A^{i'}}\tenrl{^A_{Bkl}}\tenth{^B_{j'}}$. As a connection in $\bund$ we choose $\connbund=\partial^{\lbund} \otimes 1+1 \otimes \nabla$.
These data encode the $*$-product $*_P$.

We are going to make use of the following two endomorphisms of $\lbund \otimes T\sympman$: $\tenmodrl{}$, i.e. rescaled\footnote{The function $v$ modifying $\tenrl{}$ is taken with respect to the metric $g_{ab}$ induced by the tetrad. Obviously $\vol_M$ and the volume form given by the determinant of $\theta$ coincide in such case.} by $v$ version of  $\tenrl{^A_B^{a'}_{b'}}$ (defined by the curvature of $\partial^{\lbund}$ and the tetrad which raises index ${}^{a'}$), and $\tent{}$ given by $\tent{^A_B^{a'}_{b'}}=\tenth{^{Aa'}}\tenth{_{Bb'}}$. 
As a starting point one may take the following version of Palatini action
\begin{equation*}
\mathcal{S}_{P}=\int_{\sympman}\Tr\tenmodrl{}\Theta\,\frac{\omega^n}{n!}
\end{equation*}
The deformation procedure yields particularly simple expression due to $\partial_i \tent{}=0$, $\Tr(\curvbund_{a b} \{\tenrl{},\tent{}\})=0$ and $\Tr(\curvbund_{a b}\curvbund_{c d} \{\tenrl{},\tent{}\})=0$.
\begin{equation}
\label{pal_def_act_h2}
\widehat{\mathcal{S}}_{P}=\tr_{*_P}(\tenmodrl{}*\tent{})=\int_{\sympman}\Tr\Big(\tenrl{}\tent{} +h^2 s_2\tenrl{}\tent{} +O(h^3)\Big)\vol_M
\end{equation}
The variation with respect to $\delta\tenth{}$ leads to the equations
\begin{subequations}
\begin{equation}
(1+h^2 s_2)\left(R_{ab}-\frac{1}{2}g_{ab}R\right)+O(h^3)=0
\end{equation}
clearly equivalent (up to $h^2$) to the condition $R_{ab}=0$. The variation of the connection field $\delta\undertilde{\Gamma}$ produces
\footnote{The variation gives 
\begin{equation*}
w(\tensor{\Gamma}{^c_{ab}}-\tensor{\Gamma}{^c_{ba}})=\\
\tensor{\delta}{^c_a}\left(\frac{\partial w}{\partial x^b} - \tensor{\Gamma}{^d_{bd}}w \right)
-
\tensor{\delta}{^c_b}\left(\frac{\partial w}{\partial x^a} - \tensor{\Gamma}{^d_{ad}}w \right)
\end{equation*}
with the tensor density $w=\sqrt{-g}(1+h^2s_2)$. Contraction of this relation enables expressing $\tensor{\Gamma}{^d_{ad}}$ in terms of $\tensor{\Gamma}{^d_{da}}$, leading in turn to (\ref{palatinitorsion}).
}
\begin{equation}
\label{palatinitorsion}
(1+h^2 s_2)\tentor{^a_{bc}}=\frac{h^2}{n-1} \tensor{\delta}{^a_{[b}} \frac{\partial s_2}{\partial x^{c]}}+O(h^3)
\end{equation}
\end{subequations}
where $\tentor{^a_{bc}}=\tensor{\Gamma}{^a_{cb}}-\tensor{\Gamma}{^a_{bc}}$ is the torsion tensor of the connection $\nabla$. Thus, one obtains the theory with vanishing Ricci tensor and nonvanishing torsion generated by the scalar $s_2$. A quick calculation shows that the trace-free part of $\tentor{^c_{ab}}$ is equal to zero. Equation (\ref{palatinitorsion}) means that for 
\begin{equation*}
\tentor{^a_{bc}}=\ttorzero{^a_{bc}}+h\ttorone{^a_{bc}}+h^2\,\ttortwo{^a_{bc}}+\dots
\end{equation*}
one has $\ttorzero{^a_{bc}}=\ttorone{^a_{bc}}=0$ and
\begin{equation*}
\ttortwo{^a_{bc}}=\frac{1}{n-1} \tensor{\delta}{^a_{[b}} \frac{\partial s_2}{\partial x^{c]}}.
\end{equation*}
Connection coefficients for $\nabla$ are given by
\begin{equation*}
\teng{^a_{bc}}=\frac{1}{2} g^{ak} \left(\frac{\partial g_{bk}}{\partial x^c}+
\frac{\partial g_{ck}}{\partial x^b}-\frac{\partial g_{bc}}{\partial x^j}+
\tentor{_b_k_c}+\tentor{_c_k_b}-\tentor{_k_b_c}
\right).
\end{equation*}
Hence, $\tgzero{^a_{bc}}$ and $\tgone{^a_{bc}}$ are still expressed by relations (\ref{eh_gamma_zero}) and (\ref{eh_gamma_one}). For $\tgtwo{^a_{bc}}$ one computes that
\begin{subequations}
\begin{equation}
\label{pal_gam2_corr}
\tgtwo{^a_{bc}} = \frac{1}{2}\zeroed{g}^{\;ak}\left(\zeroed{\nabla}_{c} \twoed{g}_{kb} + \zeroed{\nabla}_{b} \twoed{g}_{kc} - \zeroed{\nabla}_{k} \twoed{g}_{bc}\right)
- \zeroed{g}^{\;ak\;} \oned{g}_{kl} \tgone{^{\,l}_{bc}}
+\frac{1}{2(n-1)}\left(
\tensor{\delta}{^a_c} \frac{\partial s_2}{\partial x^b }-
\zeroed{g}_{bc} \zeroed{g}^{\;ak} \frac{\partial s_2}{\partial x^k }
 \right).
\end{equation}
Corrections to Riemann tensor are again given by (\ref{eh_riemann_corr}). Substituting them to $R_{ab}=0$ we obtain that for $h^0$ and $h^1$  relations (\ref{eh_g0_corr}) and (\ref{eh_g1_corr}) remain valid. However, equations for $\twoed{g}_{ab}$ take the following form
\begin{multline}
\label{pal_g2_corr}
\zeroed{g}^{\;kl}\left(\zeroed{\nabla}_k \zeroed{\nabla}_a \twoed{g}_{bl} + \zeroed{\nabla}_k \zeroed{\nabla}_b \twoed{g}_{al} - \zeroed{\nabla}_k \zeroed{\nabla}_l \twoed{g}_{ab} - \zeroed{\nabla}_a \zeroed{\nabla}_b \twoed{g}_{kl} 
\right)=\\
=2 \zeroed{\nabla}_a \zeroed{\nabla}_b s_2 + \frac{1}{n-1} \, \zeroed{g}_{ab} \zeroed{g}^{\;kl}\zeroed{\nabla}_k \zeroed{\nabla}_l s_2
-4\tgone{^k_{l[k}}\tgone{^l_{b]a}}
+4\,\zeroed{g}^{\;rk} \zeroed{\nabla}_{[r} \left(\,\tgone{^l_{b]a}} \oned{g}_{kl}\right).
\end{multline}
\end{subequations}
Like in the case of Einstein-Hilbert action, one can easily guess special solution of (\ref{pal_g2_corr}) by requiring that $\oned{g}_{ab}=0$, i.e. that classical first order perturbation vanish. It reads
\begin{equation}
\label{pal_spec_sol}
\twoed{g}_{ab}=- \frac{1}{n-1}s_2\zeroed{g}_{ab}
\end{equation} 
For such case the correction $\tgone{^a_{bc}}$ is equal to zero, and $\tgtwo{^a_{bc}}$ is given by
\begin{equation}
\label{pal_spec_sol_conn}
\tgtwo{^a_{bc}}=- \frac{1}{2(n-1)}\tensor{\delta}{^a_b} \frac{\partial s_2}{\partial x^c}.
\end{equation} 
Repeating arguments of the previous section, one can point out that arbitrary other solution of (\ref{pal_g2_corr}), with $\oned{g}_{ab}=0$, differs from (\ref{pal_spec_sol}) by a classical perturbation of $\zeroed{g}_{ab}$. Let us observe that since deformation of the action given by (\ref{pal_def_act_h2}) depends up to $h^2$ solely on $s_2$, then  corrections (\ref{pal_spec_sol}), (\ref{pal_spec_sol_conn}) are also expressible in terms of $s_2$. Now, $s_2$ is related to the curvature of $\connsymp$  by the formula (\ref{pal_def_act_h2}). In particular this means, that for the Moyal case of flat $\connsymp$, one is dealing with undeformed theory even at $h^2$. 

\section{Relation to the theory of Seiberg-Witten map}
Let us explain how proposed models can be understood in terms of the theory of Seiberg-Witten map. This becomes quite straightforward when one combines results of \cite{dobrski2} with the property (\ref{traceinvar}). Indeed, what \cite{dobrski2} states is that Seiberg-Witten map is an local isomorphism of $*$-product algebras, while the relation (\ref{traceinvar}) says that the trace functional is invariant on such isomorphisms. 

More precisely, suppose that one prescribes to each frame $e$ in $\bund$ $*$-product isomorphism $M_{\langle e \rangle}$ which transforms the initial global product $*$ to the local one $*_S$. (Recall that $*_S$ is nothing but matrix multiplication with commutative product of entries replaced by Fedosov $*$-product of functions). Thus
\begin{equation*}
M_{\langle e \rangle}(F_{\langle e \rangle}*G_{\langle e \rangle})=M_{\langle e \rangle}(F_{\langle e \rangle})*_S M_{\langle e \rangle}(G_{\langle e \rangle})
\end{equation*}
where $F_{\langle e \rangle}$, $G_{\langle e \rangle}$ are matrices representing endomorphisms $F$ and $G$ in the frame $e$. It turns out (\cite{dobrski2} section 3.1) that if we switch to different frame $\widetilde{e}=eg^{-1}$ then $M_{\langle e \rangle}$ and $M_{\langle \widetilde{e} \rangle}$ are related by
\begin{equation}
\label{seibwitt_covar}
M_{\langle \widetilde{e} \rangle}(F_{\langle \widetilde{e} \rangle})=
\widehat{g}_{\frmbr{e}}(g,\gambund) *_S
M_{\langle e \rangle}(F_{\langle e \rangle}) *_S
\widehat{g}_{\frmbr{e}}^{-1}(g,\gambund)
\end{equation}
with $\widehat{g}_{\frmbr{e}}(g,\gambund)=g+O(h)$ dependent both on $g$ and connection one-forms $\gambund_i$ in the frame $e$ and their derivatives. Moreover, if we combine two gauge transformations, then $\widehat{g}$ fulfills ``consistency conditions'' (compare \cite{jurco0, schupp}) given by
\begin{equation}
\label{triv_g_consistency}
\widehat{g}_{\frmbr{e}}(g'g,\gambund)=\widehat{g}_{\frmbr{\tilde{e}}}(g',g\gambund g^{-1}+g \rmd g^{-1})*_S \widehat{g}_{\frmbr{e}}(g,\gambund).
\end{equation}
Thus, $M$ and $\widehat{g}$ behave exactly like Seiberg-Witten map \cite{seibwitt}. Indeed, if $M$ is set up with Fedosov's techniques of generating $*$-product isomorphisms, then one can compute\footnote{Fedosov construction enables \emph{computation} of Seiberg-Witten map, up to arbitrary order in $h$, by its recursive techniques. This situation is rather different from the usual framework, where the Seiberg-Witten equations must be \emph{solved}.} $M$ and $\widehat{g}$, and for the case of $*_S$ given by Moyal product $*_T$  obtain results which are well-known expressions for Seiberg-Witten map (\cite{dobrski2} section 4).


We are going to rewrite investigated actions in terms of Seiberg-Witten map. Let $M(F)=\widehat{F}$, as it is justified by relations (\ref{triv_g_consistency}) and (\ref{seibwitt_covar}). Also, let us separately distinguish Moyal case of $*_S=*_T$, for which $\connsymp$ is flat,  one works in Darboux coordinates with coefficients of $\connsymp$ equal to zero, and the trace functional $\tr_{*_T}$ is given by the integral (\ref{tracemoyal}). Suppose that supports of endomorphisms under consideration are small enough to be covered by a single frame in $\bund$, and -- in Moyal case -- by a single Darboux coordinates. Then, due to property (\ref{traceinvar}), actions considered in this paper can be locally rewritten as follows.
\begin{center}
\begin{tabular}{r| l |l}
& arbitrary $*_S$ & $*_S=*_T$\\
\hline
\rule{0pt}{4ex}
$\widehat{\mathcal{S}}_{EH_{1A}}=$
&
$\displaystyle \tr_{*_S}(\,\widehat{\modricciend\,})$
&
$\displaystyle \int_{\mathbb{R}^{2n}}\Tr(\,\widehat{\modricciend\,}) \; \rmd^{2n}x$\\
\rule{0pt}{4ex}
$\widehat{\mathcal{S}}_{EH_{1B}}=$
&
$\displaystyle \tr_{*_S}(\widehat{\ricciend\,} *_S \widehat{V})$
&
$\displaystyle \int_{\mathbb{R}^{2n}}\Tr(\widehat{\ricciend\,} *_T \widehat{V}) \; \rmd^{2n}x$\\
\rule{0pt}{4ex}
$\widehat{\mathcal{S}}_{EH_{2A}}=$
&
$\displaystyle \tr_{*_S}(\,\widehat{\modriemannend\,})$
&
$\displaystyle \int_{\mathbb{R}^{2n}}\Tr(\,\widehat{\modriemannend\,}) \; \rmd^{2n}x$\\
\rule{0pt}{4ex}
$\widehat{\mathcal{S}}_{EH_{2B}}=$
&
$\displaystyle \tr_{*_S}(\widehat{\riemannend\,} *_S \widehat{V})$
&
$\displaystyle \int_{\mathbb{R}^{2n}}\Tr(\widehat{\riemannend\,} *_T \widehat{V}) \;\rmd^{2n}x$\\
\rule{0pt}{4ex}
$\widehat{\mathcal{S}}_{P}=$
&
$\displaystyle \tr_{*_S}(\widehat{\tenmodrl{}\,} *_S \widehat{\Theta})$
&
$\displaystyle \int_{\mathbb{R}^{2n}}\Tr(\widehat{\tenmodrl{}\,} *_T \widehat{\Theta})\; \rmd^{2n}x$
\end{tabular}
\end{center}
Let us observe that such setting clarifies how considered models are related to the spacetime noncommutativity described by $*_S$. Indeed, due to (\ref{tracecommutation}), above mentioned local versions of action functionals are invariant with respect to gauge transformations (\ref{seibwitt_covar}) realized by means of $*_S$. Thus, one is able to reasonably claim that models considered in this paper correspond to noncommutativity of spacetime generated by Fedosov product of functions $*_S$.

\section{Discussion}
We have obtained number of nonequivalent geometric deformations of vacuum Einstein relativity. They have been analyzed at $h^2$ order, starting from the action functional, through field equations, up to corrections to the metric which have been explicitly given for the case of $\widehat{\mathcal{S}}_{EH_{1A}}$, $\widehat{\mathcal{S}}_{EH_{1B}}$ and $\widehat{\mathcal{S}}_{P}$. Using results of \cite{dobrski2}, we have pointed out the relation between proposed models, the theory of Seiberg-Witten map, and the noncommutativity of the spacetime described by Fedosov $*$-product generated by symplectic form $\omega$ and symplectic connection $\connsymp$.

The construction scheme we have adopted, relies on the geometric deformation of product of endomorphisms, but it does not include deformations of other geometric data like connection, tensor product, exterior algebra of forms or contraction operator. (Approaches aiming at modifying various structures of classical geometry in the deformation quantization framework certainly exist. These are e.g. \cite{zumino,chaichian2,vassilevich}). The advantage of our approach consists in immediate interpretation in terms of Seiberg-Witten map. On the other hand, the price is that the noncommutativity does not appear as a fundamental structure modifying all the geometry, but rather may seem to be a kind of ``extra interaction'' entering to action functionals via the procedure described in the introduction. 

The multiplicity of models arises as a consequence of ambiguity in translating traditional action functionals to the language of traces of endomorphisms of some bundle. From the gauge simplicity point of view, actions $\widehat{\mathcal{S}}_{EH_{1A}}$ and $\widehat{\mathcal{S}}_{EH_{1B}}$ seem to be most straightforward as they correspond to the natural $GL(2n,\mathbb{R})$ gauging. On the other hand, action $\widehat{\mathcal{S}}_{P}$ produces especially simple expressions for deformed field equations and for corrections to the metric. 

The remarkable problem related to presented models concerns incompatibility of the volume forms -- metric and symplectic ones. Both proposed solutions (rescaling one of the endomorphisms and multiplication by $V$) seem to be a bit unnatural. One can suspect that this problem is related to fixing symplectic structure as a nondynamical background. Notice however, that Fedosov construction provides natural framework for the variation of the symplectic data. Moreover, it could turn out that some refinements to the Fedosov theory should be made, to put the metric into the internal structure of the deformation quantization procedure. Such considerations are hoped to be covered in author's subsequent work. 

Let us briefly discuss diffeomorphism invariance of proposed models. Clearly they are diffeomorphism invariant in the passive sense, since all actions, field equations and derived corrections to the metrics are given in either explicitly global or coordinate covariant manner. However, they are not invariant under active diffeomorphisms. Again, this issue originates in fixing symplectic data as a nondynamical background. Such observation is a further argument for considering dynamics of $\omega$ and $\connsymp$ as a natural next step within Fedosov formalism.

Due to the symmetries of the Riemann tensor, in all considered cases imaginary terms at $h^1$ have vanished. It must be stressed however, that we have no clear evidence that the same stays true for other odd powers of $h$. Thus, some further analysis of the reality of proposed actions should be performed. This suggests deeper investigation of the structure of the trace functional, which seems to be rather hard task (but not hopless, as it can inferred from Fedosov's results \cite{fedosov_on_the_trace,fedosov_atiyah_bott_patodi} on relating $\tr_*(1)$ to integrals of characteristic classes of $T\sympman$ and $\ebund$).  On the other hand, construction of some appropriate involution operator in the Fedosov algebra may be useful and it is also matter of author's further interest. 

Finally one could be interested, how the present work is related to the well known existence of \emph{closed} $*$-products (compare e.g. \cite{felder}). First of all, the existence of such 
products has been investigated for functions but not for endomorphisms (to the best of author's knowledge). Moreover, if one is going to treat Seiberg-Witten map in more or less fundamental manner, then nontriviality of the trace is what should be expected. Indeed, as it was argumented in the previous section, the nontrivial trace could be interpreted as the object carrying information about the globalization of Seiberg-Witten map.

\section*{Acknowledgments}
I am very grateful to professor Maciej Przanowski for his thorough interest in my work and many helpful remarks. I am also indebted to professors Piotr Kosi\'nski and Anatol Odzijewicz for a number of valuable comments.

\end{document}